\documentstyle[epsfig,12pt]{article}
\input{epsf.sty}
\begin{document}
\parskip 0.3cm
\begin{titlepage}
\begin{flushright}
ETH-TH/00-07  \\
July 2000 
\end{flushright}

\begin{centering}
{\Large \bf  Inclusive and Semi-Inclusive Polarized DIS Data Revisited }\\

 \vspace{.9cm}
{\bf  D. de Florian\footnote{Partially supported 
by the EU Fourth Framework Programme ``Training and Mobility of Researchers'', 
Network ``Quantum Chromodynamics and the Deep Structure of
Elementary Particles'', contract FMRX--CT98--0194 (DG 12 -- MIHT).} }
\vspace{.05in}
\\ {\it Theoretical Physics, ETH Z\"urich CH-8093, Switzerland}\\
e-mail: dflorian@itp.phys.ethz.ch \\
 \vspace{.4cm}
{\bf R. Sassot\footnote{Partially supported by 
CONICET, ANPCyT, and Fundaci\'on Antorchas.}} \\
\vspace{.05in}
{\it  Departamento de F\'{\i}sica, 
Universidad de Buenos Aires \\ 
Ciudad Universitaria, Pab.1 
(1428) Bs.As., Argentina} \\
e-mail: sassot@df.uba.ar \\
\vspace{1.5cm}
{ \bf Abstract}
\\ 
\bigskip
\end{centering}
{\small
We perform a combined next to leading order analysis to both inclusive
and semi-inclusive polarized deep inelastic scattering data, assessing the impact of new data in the extraction of the sea
quark and gluon spin dependent densities.  In particular, we find that
semi-inclusive data show a clear preference for a positive $\Delta \overline{u}$ distribution, although are almost insensitive to different alternatives to $\Delta \overline{d}$.}

\vspace{0.2in}
\vfill

\end{titlepage}
\vfill\eject

\noindent{\Large \bf 1 \,\, Introduction}\\

Over the last decade, there has been a remarkable improvement in our knowledge of the spin structure of the nucleon at partonic level, driven both by the increase in precision and volume of data on spin dependent structure functions,
and also on theoretical and phenomenological work around the ever increasing
wealth of experimental evidence  \cite{rev}. In the process, the extraction of spin dependent parton densities has evolved from mere extensions of simple scale independent SU(6)-inspired models, to consistent next to leading order (NLO) QCD global analysis based on a much more reduced set of model dependent hypotheses. 

In spite of this significant progress, until very recently it was almost
unavoidable to make rather crude assumptions about the way in which sea quarks
are polarized, or more precisely, how polarization is distributed among the
different sea quark flavours.  The main reasons for this being the inability
of totally inclusive data to discriminate between quarks and antiquarks, and
the lack of enough statistics in the pioneering polarized semi-inclusive
experiments.  Because of these circumstances, up to now even the most refined
global analyses have simply assumed polarization to be evenly distributed
among the sea flavours, or in the most ambitious attempts, they have tried to
discriminate between light and heavier flavours, but always in compliance with
SU(2) isospin symmetry assumptions \cite{global1,DSS,global2,ghosh}.  

Motivated by the growing evidence supporting an SU(2) symmetry breaking between unpolarized sea quark distributions, more recently there has been a great deal of activity concerning sea quarks polarization, including theoretical and phenomenological models advocating different degrees and realizations of SU(2) symmetry breakdowns \cite{mod}, and even attempts to find evidence for them in the data \cite{mori}.

Semi-inclusive asymmetries are sensitive to contributions coming from specific combinations of quarks of different flavour and nature, so in principle they can
shed some light on the sea quark polarization issue. The theoretical framework needed to deal with them consistently up to NLO accuracy is well known \cite{NPB}, and indeed this kind of data was successfully included in a NLO global analysis some time ago \cite{DSS}. However, the statistical weight of the data available at that time \cite{SMC}
hindered a significant impact on the parton distribution extraction. Since then, the Hermes Collaboration \cite{Hermes} have presented new semi-inclusive data with
greater precision, circumstance that justifies a new analysis taking into account also the new additions to the inclusive data set. 

In this paper we perform a combined next to leading order analysis of both inclusive and semi-inclusive polarized deep inelastic scattering data, releasing the usual assumptions about flavour symmetry relations between the polarization of both valence and sea quarks, and we assess the role of semi-inclusive data in the experimental evaluation of these relations.

Performing this analysis we have found that semi-inclusive data constitute a
clear constraint on the polarization of `up' antiquarks, information which combined with that coming from inclusive measurements, fix the polarization of 
valence quarks of the same flavour. However, `down' antiquarks polarization remains loosely bounded allowing a large uncertainty on `down' valence quark densities. As in previous analyses, the data imply a weak constraint on the 
polarized gluon density.

This paper is organized as follows: In section 2 we review the framework for
global analyses of inclusive and semi-inclusive data, and present our
conventions. In sections 3 and 4 we present the result for the fits to
inclusive and combined (inclusive {\it plus} semi-inclusive) data,
respectively. Finally, in section 5, we give our conclusions and in the appendix
we present the parameters for the fits.\\

\noindent{\Large \bf 2 \,\, Conventions and Framework}\\

Throughout the present analysis we adopt the $\overline{MS}$ factorization
scheme, as implemented in our previous analysis \cite{DSS}, for writing inclusive and semi-inclusive asymmetries in terms of parton distributions up to NLO accuracy. For the inclusive asymmetries we take as usual
\begin{equation}
\label{eq1}
 A_{1}^{N}(x,Q^2) \simeq
\frac{g_{1}^{N}(x,Q^2)}{F_{1}^N(x,Q^2)},
\end{equation}
where the spin dependent structure function is given by
\begin{eqnarray}
\label{eqg1}
 g_{1}^{N}(x,Q^2)=\frac{1}{2} \sum_{q, \bar q} e_q^2
\left \{ \Delta q (x,Q^2)   + \frac{\alpha_s(Q^2)}{2\pi} [ \Delta C_q \otimes
\Delta q  +  \Delta C_g \otimes
\Delta g ] \right\} ,
\end{eqnarray}
and in the case of three flavours can be alternatively written as
\begin{eqnarray}
\label{eqg1new}
g_1^N(x,Q^2) &=&  \left( \pm \frac{1}{12} \Delta q_3^{NS} + \frac{1}{36} \Delta q_8^{NS} 
+ \frac{1}{9} \Delta \Sigma \right) \otimes \left(1+ \frac{\alpha_s}{2\pi} \Delta C_q
\right) \nonumber\\
&+& \sum_q e_q^2  \frac{\alpha_s}{2\pi} \Delta g \otimes  \Delta C_g.
\end{eqnarray}
In equation (3) the $\pm$ sign corresponds to either proton ($N=p$) or neutron  ($N=n$) targets and 
\begin{eqnarray}
\label{eq38sigma}
\Delta q_3^{NS} &\equiv & \left( \Delta u +   \Delta \bar{u} \right) - \left( \Delta
  d +   \Delta \bar{d} \right) \nonumber \\
\Delta q_8^{NS} &\equiv& \left( \Delta u +   \Delta \bar{u} \right) + \left( \Delta
  d +   \Delta \bar{d} \right) -2 \left( \Delta s +   \Delta \bar{s} \right) \nonumber \\
\Delta \Sigma &\equiv& \left( \Delta u +   \Delta \bar{u} \right) + \left( \Delta
  d +   \Delta \bar{d} \right) + \left( \Delta s +   \Delta \bar{s} \right) 
\, .
\end{eqnarray}
The first two (non-singlet) distributions evolve in $Q^2$ independently, 
whereas the gluon and the singlet distributions are coupled in the evolution.

It is worth noticing that whereas $\Delta q_3^{NS}$ can be obtained directly 
from data on the difference between the proton and the neutron spin dependent structure function $g_1^p-g_1^n$, the three remaining distributions that appear in
equation (\ref{eqg1new}) have to be obtained somewhat indirectly, i.e. exploiting their different scale dependence. Unfortunately, the limited amount of data available, and the rather restricted range in $x$ and $Q^2$ covered, do not allow to do a precise determination of all the distributions. Nevertheless it should be kept in mind that even with an unbounded set of inclusive data of unlimited precision, it would only be possible to determine combinations of $\Delta q + \Delta \bar{q} = \Delta q_v + 2 \Delta \bar{q}$ (with $q=u,d,s$), but
{\it never}  $\Delta q_v$ and $\Delta \bar{q}$ separately, as can be seen from equation (\ref{eq38sigma}); therefore, the data is insensitive to
the symmetry breaking in the sea sector.\footnote{In
  this sense, claims  stating that 
 `the data tolerate strong flavor symmetry violations in the sea sector' (as
 in reference \cite{ghosh}) are
  just  rather void statements, since of course  it is always possible to modify
  $\Delta \bar{q}$ in such a way that $\Delta q + \Delta \bar{q}$ remains      
  unchanged.}

To be able to determine both $\Delta q_v$ and $\Delta \bar{q}$ one
has to introduce some other independent input in the analysis, such as weak
structure functions and Drell-Yan production, for which the theoretical framework is well known \cite{alpha,drellyan} but unfortunately there is no data yet, or semi-inclusive deep inelastic scattering (SIDIS) asymmetries. 

In order to perform the extraction of polarized valence and sea quark densities, we include SIDIS asymmetries measured by Hermes and SMC, which are given by
\begin{equation}
\label{eq2}
 \left.
A_{1}^{N\,h}(x,Q^2) \right|_Z \simeq \frac{\int_{Z} dz\,
g_{1}^{N\,h}(x,z,Q^2)}{\int_{Z} dz\,F_{1}^{N\,h}(x,z,Q^2)}.
\end{equation} 
There, the semi-inclusive analog of $g_1^N$ is given by
\begin{eqnarray} 
\label{eqg1si}
g_{1}^{N\,h}(x,z,Q^2) = \sum_{q,\overline{q}} e^2_i \left
\{ \Delta q_i (x,Q^2)  D^h_{q_i}(z,Q^2)  \right.  \nonumber \\
+ \left. \frac{\alpha_s(Q^2)}{2\pi} [ \Delta q_i
 \otimes \Delta C_{ij} \otimes D^h_{q_j}
+  \Delta
q_i  \otimes \Delta C_{ig} \otimes D^h_g+ \Delta g  \otimes
\Delta C_{gj} \otimes D^h_{q_j} ] \right. \nonumber \\ 
+ \left.  \Delta M_{q_i}^h (x,z,Q^2)
+\frac{\alpha_s(Q^2)}{2\pi}[\Delta M_{q_i}^h \otimes \Delta C_{i}+\Delta
M_{g}^h \otimes \Delta C_{g}] \right \}  ,
\end{eqnarray} 
and $Z$ denotes the kinematical region covered by final state hadrons
($z_h=P\cdot h/P\cdot q>0.2$ in \cite{SMC,Hermes}). The first line in equation (\ref{eqg1si}) represents the familiar LO contribution to semi-inclusive processes, the second accounts for current fragmentation up to NLO, and the third is associated with target fragmentation and can be neglected for the kinematical region explored 
by Hermes and SMC \cite{PLB}. The full expressions for the coefficients in
equations (2) and (6) in the $\overline{MS}$ can be found in \cite{DSS} and
references therein. 
  
SIDIS data are available for different kind of targets and species of hadrons identified in the final state
($p,D,He$ and $\pi^\pm, h^\pm$ respectively). Since fragmentation functions
depend strongly  on the flavour of the originating parton, and on the charge of the
hadron, it is possible to construct alternative combinations of parton distributions 
which are independent from the ones contributing to the inclusive asymmetry.
In this way,  in principle it possible to do a full flavour
 decomposition for the polarized parton distributions.

Fragmentation functions are obtained from fits to $e^+e^- \rightarrow h$ data
 and, similarly to what  happens in inclusive DIS, those measurements allow
 only the determination of the the combination $D_q^h+D_{\bar{q}}^h$.
In order to construct the full set of fragmentation functions needed to analyze the SIDIS
data for $h^+$ and $h^-$, one has to do some extra assumptions.
In the present analysis we take into account only  contributions from $\pi$'s and $K$'s, and for 
both mesons we define a $Q^2$  and flavour independent function $\eta_D$\footnote{Therefore the $Q^2$ dependence of the
 'favored' and 'unfavored' fragmentation functions is given, in our
 approximation,  by the one of the
 total distributions  fitted from $e^+e^-$ data}, 
which allows to separate between the `favored' and `unfavored' fragmentation
functions for each flavour and final state hadron as
\begin{eqnarray}
&& D_u^{\pi^+}=D_{\bar u}^{\pi^-} = D_d^{\pi^-}=D_{\bar d}^{\pi^+}\equiv D_1^\pi= \left( D_u^{\pi^+} + D_{\bar
u}^{\pi^+}\right) \frac{1}{1+\eta_D} \nonumber \\
&& D_u^{\pi^-}=D_{\bar u}^{\pi^+}= D_d^{\pi^+}=D_{\bar d}^{\pi^-}\equiv D_2^\pi= \left( D_u^{\pi^+} + D_{\bar
u}^{\pi^+}\right) \frac{\eta_D}{1+\eta_D} \nonumber \\
&& D_s^{\pi^-}=D_{\bar s}^{\pi^+}=D_s^{\pi^+}=D_{\bar s}^{\pi^-}\equiv
\frac{1}{2}\left(  D_s^{\pi^+}+D_{\bar s}^{\pi^+} \right)
\end{eqnarray}
and
\begin{eqnarray}
&& D_u^{K^+}=D_{\bar u}^{K^-}=D_s^{K^-}=D_{\bar s}^{K^+} \equiv D_1^K =\left( D_u^{K^+} + D_{\bar
u}^{K^+}\right) \frac{1}{1+\eta_D} \nonumber \\
&& D_u^{K^-}=D_{\bar u}^{K^+}= D_s^{K^+}=D_{\bar s}^{K^-} \equiv D_2^K=\left( D_u^{K^+} + D_{\bar
u}^{K^+}\right) \frac{\eta_D}{1+\eta_D} \nonumber \\
&& D_d^{K^-}=D_{\bar d}^{K^+}=D_d^{K^+} = D_{\bar d}^{K^-}\equiv
\frac{1}{2}\left(  D_d^{K^+}+D_{\bar d}^{K^+} \right)
\end{eqnarray}
In agreement with measurements of {\it unpolarized} SIDIS data \cite{sidisunpol},
we parameterize $\eta_D$ as
 \begin{eqnarray}
\label{eqRD}
\eta_D(z)= a \frac{1-z}{1+z} \, .
\end{eqnarray}

The usual Field \& Feynman parameterization for $\eta_D$ \cite{field} corresponds to equation (\ref{eqRD}) with
$a$  set to 1. In our case we will vary the parameter $a$ around the default value of 1, as a way
to quantify the uncertainties introduced by the assumptions on the fragmentation functions in the extracted polarized parton distributions.

Regarding the normalization of equations (\ref{eq1}) and (\ref{eq2}), is customary to write the unpolarized structure functions $F^N_1(x,Q^2)$ in terms of the more familiar $F^N_2(x,Q^2)$:
\begin{equation}
F_1(x,Q^2)= \frac{F_2(x,Q^2)}{2x} \frac{[1+\gamma^2(x,Q^2)]}{[1+R(x,Q^2)]},
\end{equation}
where $\gamma^2(x,Q^2)=4 M^2 x^2/Q^2$, $R(x,Q^2)$ is the ratio between the
longitudinal and transverse photoabsortion cross sections, and we have dropped
the nucleon target label $N$. Therefore, the asymmetry for inclusive data is
given by
\begin{equation}
\label{asym}
A_1(x,Q^2)=\frac{g_1(x,Q^2)}{F_2 (x,Q^2)}
 \frac{2x\,[1+R(x,Q^2)]}{[1+\gamma^2(x,Q^2)]}, 
\end{equation}
and similarly for the semi-inclusive one.

The information on the polarized parton distributions, whose extraction is the 
goal of the global fit, is therefore fully contained in the polarized 
structure function $g_1(x,Q^2)$, which can be reconstructed using the known
(measured or computed) values 
of $A_1(x,Q^2)$, $F_2 (x,Q^2)$ and $R(x,Q^2)$.

 Of course, the $A_1(x,Q^2)$
values are obtained from measurements,  but there is still some freedom in taking  $F_2$ and $R$, i.e.  
either from experimental data or from pQCD calculations. Since both quantities 
are strongly affected by higher twist contributions and unknown higher order
corrections in the measured range of $x$ and $Q^2$, but only a leading twist
LO/NLO expression for $g_1$ is used in equation (\ref{asym}) under the assumption that higher twist effects
may cancel when taking the ratio, in the past it has been considered
 more consistent from the  perturbation theory point of view to use the pQCD result for both $F_2$ and $R$.

On the other hand, it is worth noticing that the measured quantity in polarized DIS is $A_{||}$, related to $A_{1}$ by
\begin{equation}
A_{||}= D (A_1+\eta A_2) \sim D A_1,
\end{equation}
where the depolarization factor $D$ is obtained from the {\it measured value}
of $R$. Furthermore, it turns out that the effect of using either $R_{exp}$ or
$R_{QCD}$ can represent a difference up to  20\% (40\%) at NLO (LO)   for the
smallest values of measured $x$, and still persists, although becomes smaller,
even for values of $x$ and $Q^2$  where higher twist  effects are expected to
be small (like for SMC data). An analogous effect is expected in $F_2$, however in this case the differences are found to be smaller.

 In our analysis we have decided to use the experimental value of $R$
from reference \cite{SLACR} in order to be consistent with the extraction of
$A_1$ from $A_{||}$, but still construct $F_2$  using the pQCD
expressions that can be obtained from the unpolarized analog of equations (\ref{eqg1}) and (\ref{eqg1si}) 
and the GRV98 \cite{GRV98} set of unpolarized parton distributions. 
The choice for a perturbative $F_2$, is forced by the fact that our fit is also performed to semi-inclusive data, and thus we need consistency between the normalization of inclusive and semi-inclusive asymmetries. For the time being no
experimental values of $F_2^h$ are available for the normalization of the latter, at least for the same bins of $x$ and $Q^2$ measured by SMC and Hermes.   Regarding $R^h_{exp}$, we have just assumed that the same value applies for
both inclusive and semi-inclusive data.

Clearly, the extraction of $R^{h}$
and $F_2^{h}$ from unpolarized semi-inclusive data will be a great advantage in the future in order to perform a fully consistent analysis. Nevertheless, as we
have stated before, there are still some other uncertainties in the semi-inclusive case which are expected to hinder the ones coming from eventual differences between $R$ and $R^{h}$.

We also implement a different strategy to that used in \cite{DSS} for parameterizing and extracting parton densities. The procedure consists in two stages. The first one, in which totally inclusive data is used to fix the net 
polarization carried by each  quark flavour, but without discriminating between
quarks and antiquarks, as this distinction is beyond the reach of this kind of
data. For $u$ and $d$ quarks {\em plus} antiquarks densities at the initial
scale $Q^2_0 =0.5$ GeV$^2$ we propose
\begin{equation}
 x (\Delta q+\Delta \overline{q})  =N_{q}
\frac{x^{\alpha_q}(1-x)^{\beta_q}(1+\gamma_q\,
x^{\delta_q})}{B(\alpha_q+1,\beta_q+1)+\gamma_q \, B(\alpha_q+\delta_q+1,\beta_q+1) },
\end{equation}
while for strange quarks {\em plus} antiquarks we use
\begin{equation}
 x(\Delta s + \Delta \overline{s}) = 2 N_{s}
\frac{x^{\alpha_{{s}}}(1-x)^{\beta_{{s}}}}
{B(\alpha_{{s}}+1,\beta_{{s}}+1)},
\end{equation}
with a similar parametric form for gluons
\begin{equation}
 x\Delta g =N_{g}
\frac{x^{\alpha_{{g}}}(1-x)^{\beta_{{g}}}}
{B(\alpha_{{g}}+1,\beta_{{g}}+1)}.
\end{equation}

The first moments of the quark densities $\delta q$ ($N_q$) are often related
to the hyperon beta decay constants $F$ and $D$ through the SU(3) symmetry relations
\begin{equation} 
\delta u +\delta \overline{u}- \delta d -\delta
\overline{d}\equiv N_{u} - N_{d} = F+D \equiv 1.2573
\end{equation} 
\begin{equation} 
\delta u +\delta
\overline{u}+ \delta d +\delta \overline{d}- 2(\delta s +\delta
\overline{s}) \equiv  N_{u} + N_{d}  - 4 N_{s} = 3F-D \equiv 0.579.
\end{equation}
Under such an assumption, 
 the previous equations would strongly constrain the normalization of the
 quark densities. However, as we are interested also in testing flavour symmetry at different levels, we leave aside that strong assumption and  relax the symmetry relations introducing two parameters, $\epsilon_{Bj}$ and $\epsilon_{SU(3)}$ respectively. These parameters account quantitatively for eventual departures from  flavour symmetry considerations, including also some uncertainties on the low-$x$ behavior, and  higher order corrections,
\begin{equation} 
\label{Bj}
N_{u} - N_{d} = (F+D)(1+\epsilon_{Bj}) 
\end{equation}  
\begin{equation}
\label{SU(3)}
 N_{u} + N_{d}  - 4 N_{{s}} =(3F-D)(1+\epsilon_{SU(3)}),
\end{equation}
and we take them as a measure of the
degree of fulfillment of the Bjorken sum rule \cite{BJ} and the $SU(3)$ symmetry.

Equations (\ref{Bj}) and (\ref{SU(3)}) allow to write the normalization of the three quark flavours in terms of $N_{{s}}$, $\epsilon_{Bj}$, and $\epsilon_{SU(3)}$. Notice that no constraints have been imposed on the breaking parameters since
we expect them to be fixed by data. The remaining parameters are constrained
in such a way that positivity with respect to GRV98 parton distributions is fulfilled. This is
particularly relevant at large $x$, and since no polarized data is available
in that kinematical region, we directly fix the parameters
$\beta_u=3.2\,(3.05)$, 
$\beta_d=4.05\,(3.77)$ and $\beta_g=6\,(6)$  for the NLO (LO) sets in agreement with GRV98.
Consistently with the choice for the unpolarized parton distributions, we use the values of $\Lambda_{QCD}$ given in Ref.\cite{GRV98} to compute $\alpha_s$ at LO and NLO.

In this way, in the first stage we obtain from the inclusive data set all
the information available there without any unnecessary symmetry assumptions.
The second stage of the procedure incorporates semi-inclusive data in order to extract antiquark densities, and consequently to fix valence quark densities, relaying on the 
possibility to discriminate between quarks and antiquarks given by positive and negatively charged hadron production asymmetries.
As antiquark densities we take 
\begin{equation}
 x\Delta \overline{q} =N_{\overline{q}}
\frac{x^{\alpha_{\overline{q}}}(1-x)^{\beta_{s}}}
{B(\alpha_{\overline{q}}+1,\beta_{\overline{q}}+1)},
\end{equation}
for $\overline{u}$ and $\overline{d}$ quarks, and we assume $\overline{s}=s$
since the possibility of discrimination in the $s$ sector is beyond the
precision of the data and $\beta_{\overline{q}}=\beta_s$.

The data sets analyzed in both stages include only points with $Q^2>1$ GeV$^2$
listed in Table 1,  and totaling 137, 118, and 34 points, from proton, deuteron, and
helium targets respectively, in the inclusive stage, plus 42, 24, and 18, from
proton, deuteron, and helium targets respectively, in the second stage.  In order
to take into account correlations between inclusive and semi-inclusive
measurements, we only analyze  the inclusive data for SMC and Hermes for
`averaged' bins. In the semi-inclusive case we take into account in the fit
only the most precise data concerning the production of charged $\pm$ hadrons
(without identifying pions, kaons, or other particles individually). \\
 
{\footnotesize
\begin{center}
\begin{tabular}{|c|c|c|c|c|} \hline \hline
Collaboration & Target& Final state & \# points & Refs. \\ \hline \hline
EMC          & proton& inclusive   &    10     & \cite{EMC} \\ \hline 
SMC          & proton, deuteron & inclusive &  12, 12  & \cite{SMCi} \\ \hline 
E-143        & proton, deuteron & inclusive  &  82, 82  & \cite{E143} \\ \hline
E-155        & proton, deuteron & inclusive   &    24, 24    & \cite{E155} \\ \hline
Hermes       & proton,helium& inclusive   &    9, 9   & \cite{Hermes} \\ \hline
E-142        & helium& inclusive   &    8     & \cite{E142} \\ \hline
E-154        & helium& inclusive   &    17     & \cite{E143} \\  \hline \hline
SMC          & proton,deuteron& $h^+$, $h^-$  &  24, 24  & \cite{SMC} \\ \hline 
Hermes       & proton, helium &  $h^+$, $h^-$   &    18, 18     & \cite{Hermes} \\  
\hline \hline
\end{tabular}
\end{center}
\begin{center}
{\footnotesize {\bf Table 1}: Inclusive and Semi-inclusive Data used in the fit.}
\end{center}
}
\vspace*{3mm}

\noindent{\Large \bf 3 \,\, Inclusive Data}\\

In this section we present the results coming from the analysis of the inclusive data set. 
Taking into account that in previous analyses, even those which have used
most recent inclusive data, the gluon densities have been found to be rather
poorly constrained, 
we extract three different sets with the gluon first moment  constrained to be
lower than 0.4 and labeled as `i', bounded between 0.4 and 0.7 and denoted as
`ii', and greater that 0.7 labeled as `iii', respectively, at the initial scale. Doing this, we aim to explore different regions for gluon polarization and, at the same time, to provide different scenarios
for predictions in future experiments. As we will show, data prefer gluon distributions with first moments around those values but does not really constrain them in a precise way.
\vspace{-0.5cm}
\setlength{\unitlength}{1.mm}
\begin{figure}[!hbt]
\begin{picture}(100,90)(0,0)
\put(18,-50){\mbox{\epsfxsize11.0cm\epsffile{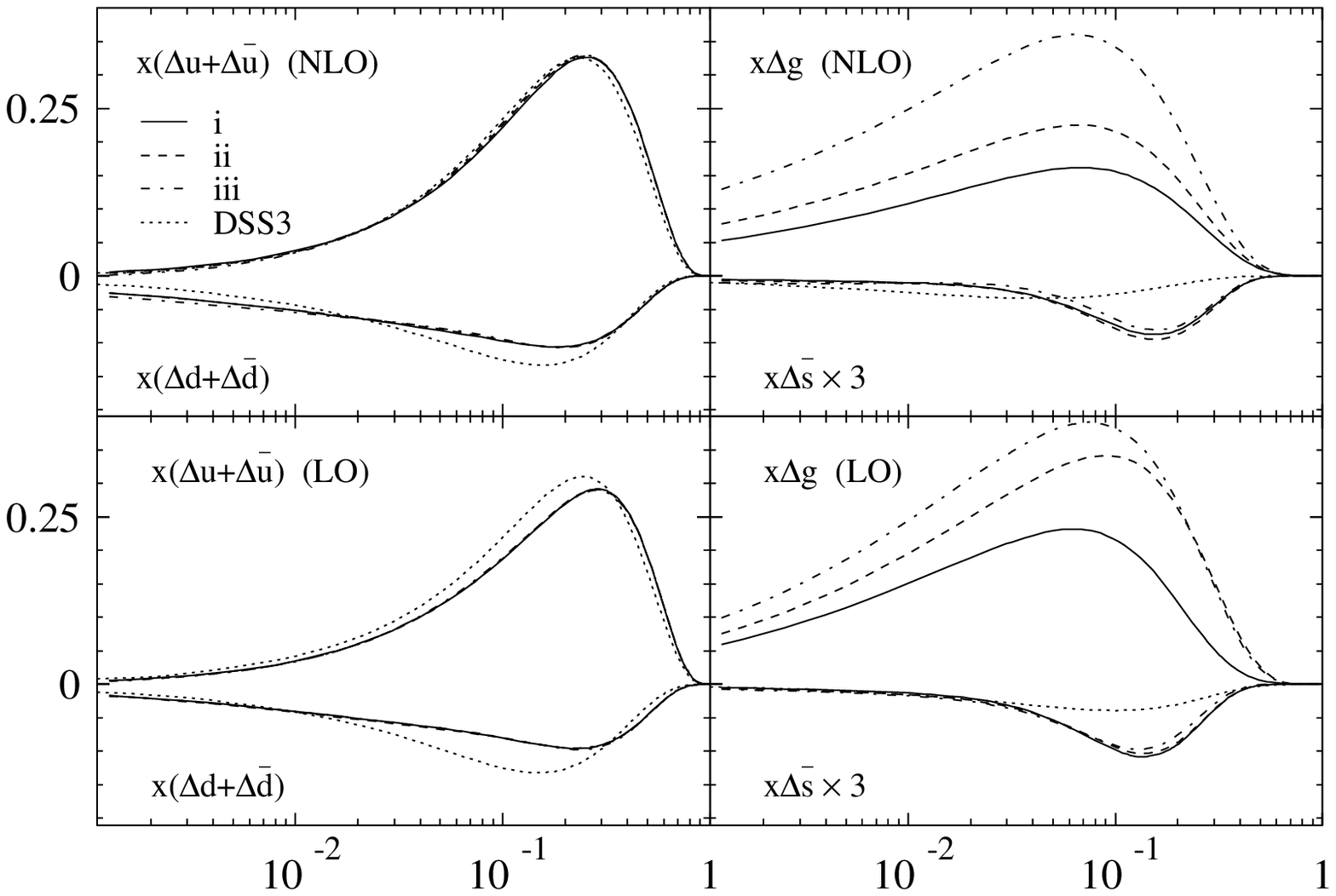}}}
\put(1,5){\mbox{\footnotesize \bf Figure 1:} {\footnotesize LO and NLO polarized parton densities from inclusive data at $Q^2=10$ GeV$^2$}} 
\end{picture}
\end{figure}
\vspace{-0.5cm}

The resulting densities from the LO and NLO analysis, and the corresponding relevant
features at $10$ GeV$^2$, are shown in Figure 1 and Table 2, respectively. The
corresponding parameters are given as an appendix (Table A3).

Even though the precision attained by this kind of fits is remarkable good, as shown in Figure 2, there seems
to be a slight systematic discrepancy between Hermes data and the results
of the remaining experiments. For these reason, in the fitting procedure we
have applied to Hermes data a normalization factor which we allow to depart
from unity up to a 5\% (it saturates at the upper limit), without almost any consequences in the obtained parton
distributions, but with a significative improvement in the $\chi^2$ values. 
\vspace{-0.4cm}
\setlength{\unitlength}{1.mm}
\begin{figure}[!hbt]
\begin{picture}(100,130)(0,0)
\put(18,-10){\mbox{\epsfxsize11.0cm\epsffile{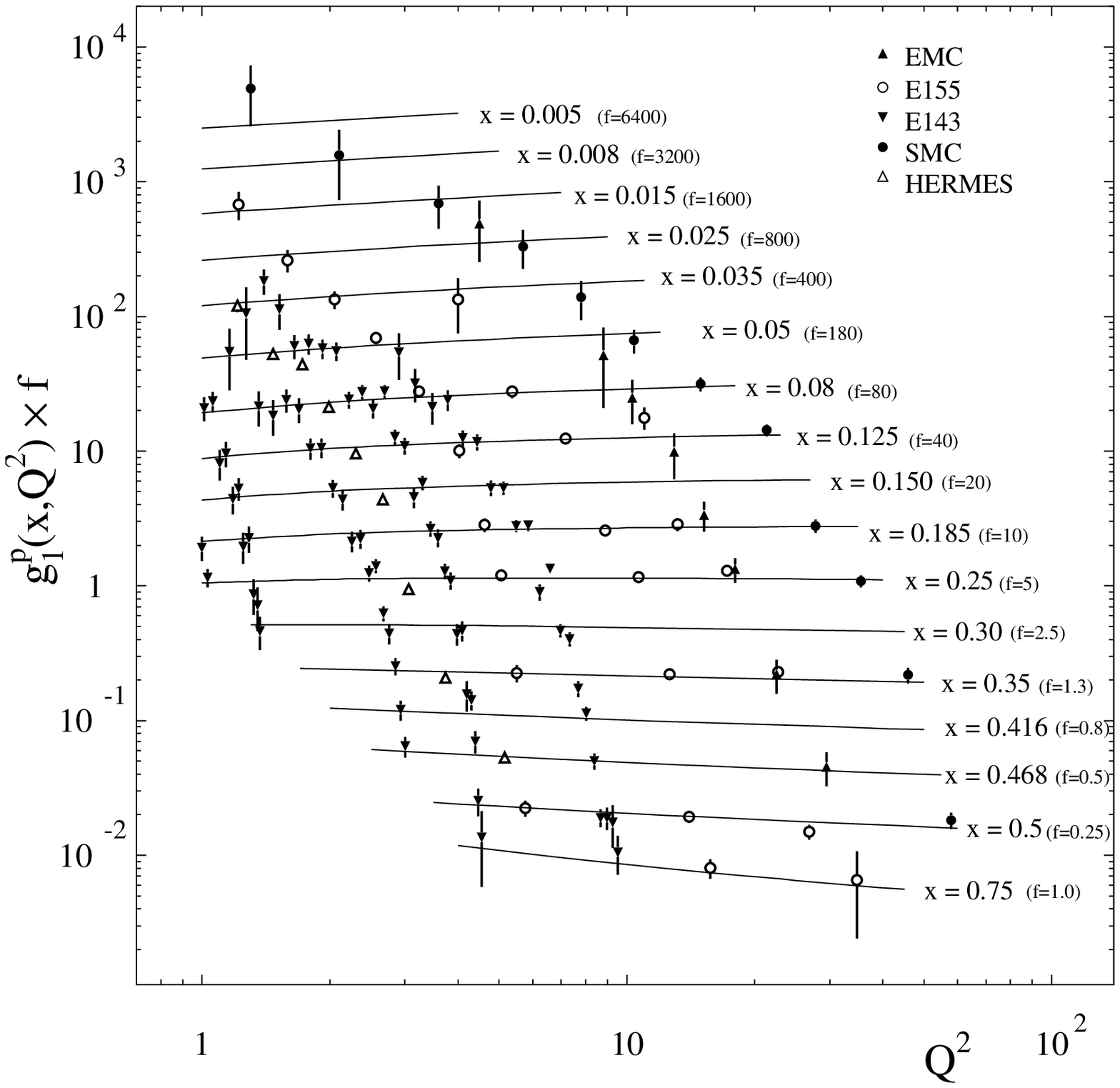}}}
\put(35,5){\mbox{\footnotesize \bf Figure 2:} {\footnotesize $g_1$ data against NLO inclusive set i)}} 
\end{picture}
\end{figure}
\vspace{-0.5cm}

As can be seen in Figure 1  the three sets have rather large
differences in the gluon distribution (with a very mild preference for set `i' at NLO and `ii' at LO), with close results for
net polarization of each quark flavour. The quality of the fits are comparable,
in fact, it is hard to distinguish between the
 inclusive asymmetries estimates coming from these sets. As it is shown in
 Table 2, the differences in $\chi^2$ are rather small, and allowing
 variations within just two units, it is possible to go from an almost
 negligible polarized gluon distribution to a highly polarized one. The reason
 for this can be easily understood from Figure 2, where all proton data for $g_1$
 is shown as a function of $Q^2$. DIS data, in principle, constrain the gluon distribution through the evolution of $g_1$, but unfortunately, available 
 data do not show a real lever-arm in $Q^2$, particularly at lower values of $x$ where the gluon density is expected to drive the evolution.

Even though  available data indicates that  $\delta g(Q^2=10)\sim 1 \pm 1$ (see Altarelli et. al. in reference \cite{global1} for a more exhaustive analysis),  we clearly have to wait for less inclusive data as in jet \cite{jet}, prompt photon \cite{pphoton}, and heavy quark \cite{hq} production to be able to determine the distribution with a reasonable precision.

In Figure 1 we have included also $u$, $d$, and $s$ distributions 
obtained in our previous analysis (DSS fit 3)\cite{DSS}. As can be observed,
the differences in the $u$ and $d$ distributions are rather small, and although there have some additions of new data in the analysis, they are mainly originated in the use of the experimental value of $R$ instead of the perturbative one as it was done in our previous analysis. This is particularly manifest in the LO case, where the perturbative approach yields $R=0$. 

It is also worth remarking that the fits prefer  $d$ distributions  steeper at small $x$ than the $u$ ones, i.e. $\alpha_d<\alpha_u$. Since there is almost no neutron data able to constrain $\Delta d+\Delta\overline{d}$ at $x<0.01$, this effect should rather be considered as a result of the extrapolation of the distributions than as  an experimental evidence of $g_1^n$ being steeper than $g_1^p$. Clearly, more precise neutron and deuteron data at small $x$ would help to clarify this feature.

Finally, 
the most prominent difference between the results of the present analysis and those from DSS, is the one found between the strange spin dependent distributions. Even though their first moments are both negative and rather close in magnitude, the shape of the distributions differ, mainly because 
$\Delta s$ is not bounded to obey the $SU(3)$ symmetric relations within the sea, constraint that was applied in the DSS analysis \cite{DSS}.
This difference give also an idea of the large uncertainties on this distribution,  which is clearly not well constrained by the available data.

Regarding the symmetry breaking for inclusive data, it is worth noticing that for the three NLO sets the Bjorken Sum Rule breaking parameter $\epsilon_{Bj}$ is found to be of the order of 1 \%, while the $SU(3)$ breaking parameter $\epsilon_{SU(3)}$ varies around a few percent, and can not be determined with good accuracy from the available data.
LO sets show similar features although with slightly higher $\chi^2$, and
larger symmetry breaking parameters. The larger value of $\epsilon_{Bj}$ at LO is mostly related to order $\alpha_s$ corrections included at NLO but absent there.

{\footnotesize
\begin{center}
\begin{tabular}{|cr|c|c|c|c|c|c|c|c|} \hline \hline
\multicolumn{2}{|c|}{set}  & $\chi ^2_I$    &
$\epsilon_{Bj}$ & $\epsilon_{SU(3)}$ & $\delta u + \delta \overline{u}$ & $\delta d+\delta \overline{d}$   & $\delta s$ & 
$\delta g$ & $\delta \Sigma$\\ \hline
& i   & 228.66  & -0.015 & -0.003 & 0.780 &-0.458 & -0.064 & 0.75 & 0.194\\ \cline{3-10} 
NLO& ii   & 228.69   & -0.018 & 0.037 & 0.776 &-0.458 & -0.071 & 1.07 & 0.175\\ \cline{3-10}
& iii   & 230.33   & -0.002 & -0.063 & 0.767 &-0.486  & -0.066 & 1.75 & 0.15 \\\hline
& i & 237.55  & -0.126 & 0.016 & 0.713 &-0.385 & -0.065 & 0.88 & 0.199\\ 
\cline{3-10}
LO & ii& 236.10  & -0.126 & 0.018 & 0.708 &-0.39 & -0.068 & 1.4 & 0.181\\ 
\cline{3-10}
& iii & 236.44  & -0.126 & 0.009 & 0.702 & -0.396  & -0.069 & 1.68 & 0.169 \\
\hline \hline
\end{tabular}
\end{center}}
\begin{center}
{\footnotesize {\bf Table 2}: First moments for LO and NLO distributions at $Q^2=10$ GeV$^2$}
\end{center}

In Table 2, we also show at $Q^2=10$ GeV$^2$ the results for the first moment of each flavour distribution and the singlet $\delta \Sigma$, which is found to be between 0.2 and 0.15 and correlated to the value of $\delta G$.

{\footnotesize
\begin{center}
\begin{tabular}{|cr|c|c|c|} \hline \hline
\multicolumn{2}{|c|}{set}  & $\Gamma^{Bj}$ & $\Gamma^{p}_1$ & $\Gamma^{n}_1$ \\ \hline
&i&  0.191 & 0.130 & -0.061 \\ \cline{3-5} 
NLO&ii&  0.191 & 0.129 & -0.062  \\ \cline{3-5}
&iii& 0.194 & 0.126 & -0.067  \\ \hline
&i  & 0.183 & 0.130 & -0.053 \\ \cline{3-5} 
LO&ii & 0.183 & 0.128 & -0.055  \\ \cline{3-5}
&iii& 0.183 & 0.126 & -0.057  \\
\hline \hline
\end{tabular}
\end{center}}
\begin{center}
{\footnotesize {\bf Table 3}: First moment of $g_1^{p-n,p,n}$ at $Q^2=10$ GeV$^2$ }
\end{center}

 Finally, in Table 3
 we present the LO and NLO results for the first moments $\Gamma_1^{p,n} =\int
 dx\, g_1^{p,n}$ and the Bjorken sum rule evaluated at $Q^2=10$ GeV$^2$. The differences in the results obtained with the three sets are mostly due to the small $x$ extrapolation, in the case of the proton, and the lack of enough precise data, for the neutron targets.\\

\newpage

\noindent{\Large \bf 4 \,\, Combined data}\\

Having obtained all the information available from the inclusive data,  we
proceed to analyze also the antiquark densities, exploiting the semi-inclusive data. The constraining power of these asymmetries can be schemed in the following way. 
The production of positively (negatively) charged hadrons   from proton targets is dominated by the convolution of $\Delta u$ ($\Delta d$) densities and `favored' fragmentation functions $D_1^\pi$, weighted with a  factor of four (one) due to the electric charge, 
\begin{eqnarray}
2 g^{\pi+(-)}_{1p} &\sim& \frac{4}{9} \Delta u \otimes D^\pi_{1(2)} + \frac{1}{9} \Delta d \otimes D^\pi_{2(1)} \nonumber \\
&&+ \frac{4}{9} \Delta \overline{u} \otimes D^\pi_{2(1)} + \frac{1}{9} \Delta \overline{d} \otimes D^\pi_{1(2)} + \frac{2}{9} \Delta \overline{s} \otimes D_3,
\end{eqnarray}
where we can neglect the suppressed  strange quark contributions, and in order to simplify the discussion,  we have taken into account only final state $\pi$'s and within a LO expression.

Factoring out the $q+\overline{q}$ distributions which are almost completely  fixed by the inclusive data we can 
write
\begin{eqnarray}
\label{eqdisc}
2  g^{\pi+(-)}_{1p} &\sim& \frac{4}{9} (\Delta u + \Delta \overline{u}) \otimes D^\pi_{1(2)}+ \frac{1}{9} (\Delta d + \Delta \overline{d}) \otimes D^\pi_{2(1)} 
 \nonumber \\
&&+ \frac{1}{9} ( \Delta \overline{d} -4 \Delta \overline{u}) \otimes(D^\pi_{1(2)}- D^\pi_{2(1)}), 
\end{eqnarray}
where it is clear that the unconstrained component of both asymmetries is proportional to the same factor $(\Delta \overline{d} -4 \Delta \overline{u})$ but with opposite signs,  and, regardless the values of $D_1$ and $D_2$,  proton asymmetries are significantly more sensitive to $\Delta \overline{u}$ due to the difference in the electric charge factors.  

Semi-inclusive asymmetries from neutron targets are proportional to a complementary combination of $\Delta \overline{d}$ and  $\Delta \overline{u}$,
and therefore are strongly dependent on $\Delta \overline{d}$. 
However, in practice, these last asymmetries have little impact in the analysis since they are heavily suppressed by the normalization coming from deuterium and helium unpolarized structure functions. 
Consequently, although $\Delta \overline{u}$ can be more easily pinned down,
  $\Delta \overline{d}$ densities are not very well constrained by present
  data. Indeed we observe this feature in the global fits: the impact of 
$\Delta \overline{d}$ in terms of $\chi^2$ is very small and a large modification in
its distribution can be easily balanced by a small modification of 
$\Delta \overline{u}$.

Taking into account the previous discussion, we parameterize $\Delta
\overline{u}$ as in equation (14) (fixing $\alpha_{\overline{u}}=
\alpha_{\overline{d}}$) but we restrict $\Delta \overline{d}$ to two somewhat extreme alternatives: one, labeled as `$+$', in which $N_{\overline{u}} = N_{\overline{d}}$, and another, labeled as `$-$', in which $N_{\overline{u}} = -N_{\overline{d}}$.
These scenarios are motivated by expectations from different models that have
been discussed in the literature. The first one ($\Delta \overline{d}=\Delta
\overline{u}$ at $Q_0^2$) corresponds to the usual assumption about SU(2) symmetry between light sea quarks. The second
alternative ($\Delta \overline{d}=-\Delta
\overline{u}$ at $Q_0^2$)   breaks the SU(2) symmetry at sea level  along the lines of the
statistical and  chiral quark soliton models \cite{mod}.

The results of allowing both the `$+$' and the `$-$' alternatives in the extended analysis that incorporates SIDIS data are presented in Table 4,
 where we show only the results associated to $\delta \overline{q}$ and $\delta
 q_v$ since those for the remaining parton densities are almost equal to the
 `inclusive' ones of Table 2. This similarity can be easily understood taking into
 account the greater precision and the amount of these data. The parameters
 for the NLO and LO fits are presented in Tables A1 and A2 in the appendix,
 respectively. 

{\footnotesize
\begin{center}
\begin{tabular}{|c|l|c|c|c|c|c|} \hline \hline
\multicolumn{2}{|r|}{set}     & $\chi ^2_T$ & $\delta u_V$ & $\delta d_V$ & $\delta \overline{u}$ & $\delta \overline{d}$ \\ \hline 
&     i$+$ &  330.52     & 0.649  &  -0.576     & 0.063 & 0.061 \\ \cline{2-7}
&     i$-$ &  330.00     & 0.686  &  -0.357     & 0.045 &-0.049 \\ \cline{2-7}
NLO&  ii$+$ &  330.98     & 0.656  &  -0.555     & 0.058 & 0.056 \\ \cline{2-7} 
&     ii$-$ &  330.22     & 0.684  &  -0.357     & 0.043 &-0.046 \\ \cline{2-7}
&     iii$+$ & 332.31     & 0.652  &  -0.583    & 0.055 & 0.053 \\ \cline{2-7} 
&     iii$-$ & 331.64     & 0.679  &  -0.392     & 0.041 &-0.044 \\ \hline
&     i$+$ &  336.24     & 0.599  &  -0.493     & 0.060 & 0.060 \\ \cline{2-7}
&     i$-$ &  335.54     & 0.629  &  -0.291     & 0.043 &-0.043 \\ \cline{2-7}
LO&   ii$+$ &  334.19     & 0.592  &  -0.498     & 0.058 & 0.058 \\ \cline{2-7} 
&     ii$-$ &  334.03     & 0.624  &  -0.295     & 0.042 &-0.042 \\ \cline{2-7}
&    iii$+$ &  334.26     & 0.591  &  -0.490     & 0.056 & 0.056 \\ \cline{3-7} 
&    iii$-$ &  333.69     & 0.624  &  -0.298     & 0.041 &-0.041 \\ 
\hline \hline
\end{tabular}
\end{center}}
\begin{center}
{\footnotesize {\bf Table 4}: $\chi ^2$ of the LO and NLO fits and the first moment of the distributions obtained from the combined analysis at $Q^2=10$ GeV$^2$.}
\end{center}

In Figure 3 we show the corresponding parton densities (only for sets i$+$ and i$-$, as the other ones are very similar), and in Figures 4 and 5, the comparison with semi-inclusive and inclusive data, respectively. The lines in Figures 4 and 5 interpolate, as usual, the results of the fit at the  measured value of $Q^2$ of each data point, and for E-143 we show only the averaged data. For the experiments that provide only the values of $g_1$, we have computed the corresponding asymmetries with the procedure discussed above.

\setlength{\unitlength}{1.mm}
\begin{figure}[!htb]
\begin{picture}(100,90)(0,0)
\put(12,-60){\mbox{\epsfxsize12.0cm\epsffile{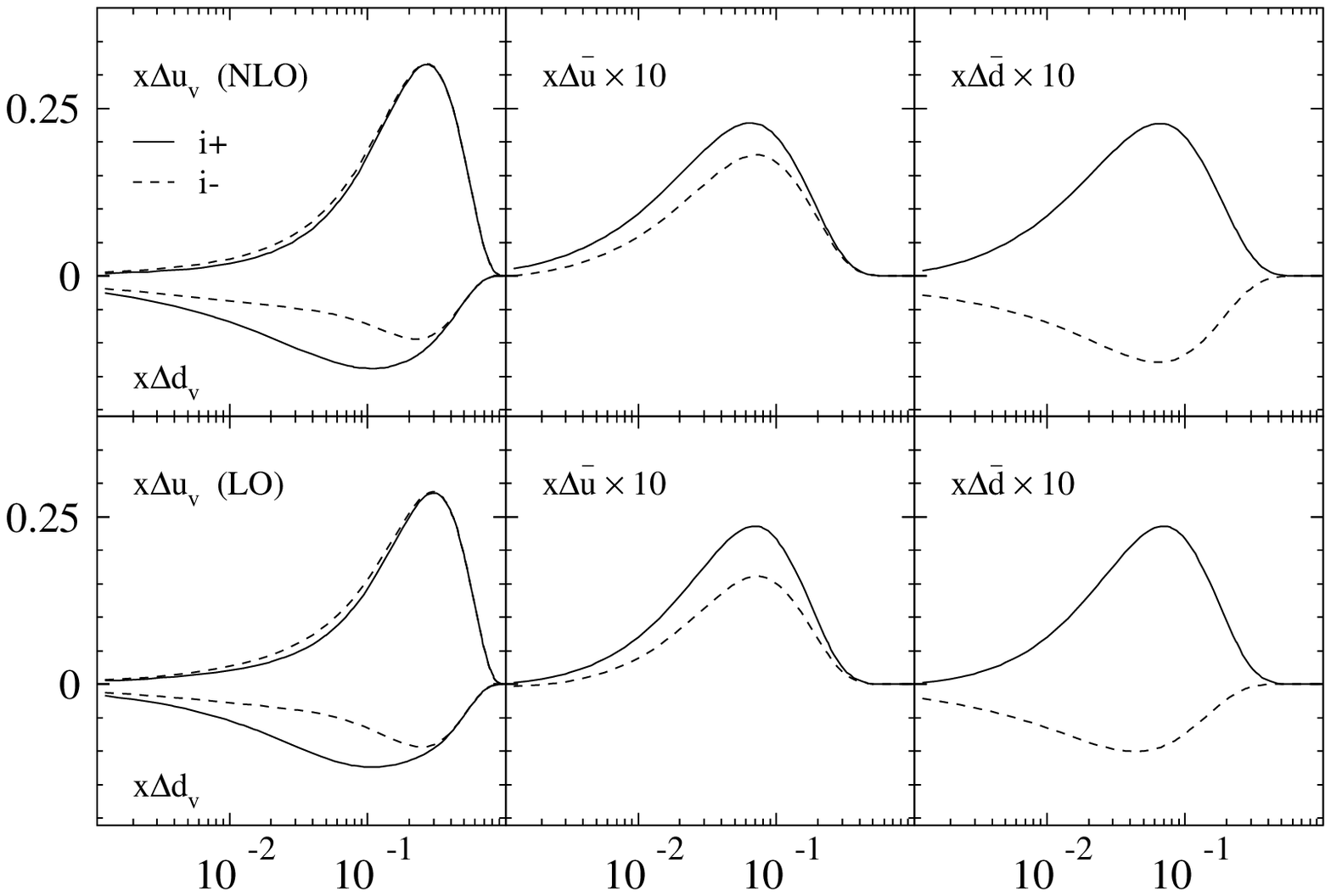}}}
\put(5,0){\mbox{\footnotesize \bf Figure 3:} {\footnotesize NLO polarized
    densities from fits to combined data at $Q^2=10$ GeV$^2$}} 
\end{picture}
\end{figure}

In Figure 4 we include also an estimate for the semi-inclusive asymmetries  computed using the parton distributions obtained in the analysis of inclusive data and assuming $SU(3)$ symmetry of the sea, i.e. $\Delta \overline{u}=\Delta \overline{d}=\Delta s$.
As it can be noticed, the differences are significative in the case of Hermes
data from proton targets, which completely dominates the fit of semi-inclusive
data, whereas  no distinction can be observed in the almost vanishing
$He$ asymmetries.

 Furthermore, as can be expected from equation (\ref{eqdisc}) the effect 
of a positive $\Delta \overline{u}$ distribution is to decrease (increase) the
asymmetry for the production of positively (negatively) charged hadrons from
proton targets with respect to the case of the $SU(3)$ symmetric result (which
of course corresponds to a negative $\Delta \overline{u}$),  as required by
the experimental data.

\setlength{\unitlength}{1.mm}
\begin{figure}[!hbt]
\begin{picture}(100,150)(0,0)
\put(10,-15){\mbox{\epsfxsize13.5cm\epsffile{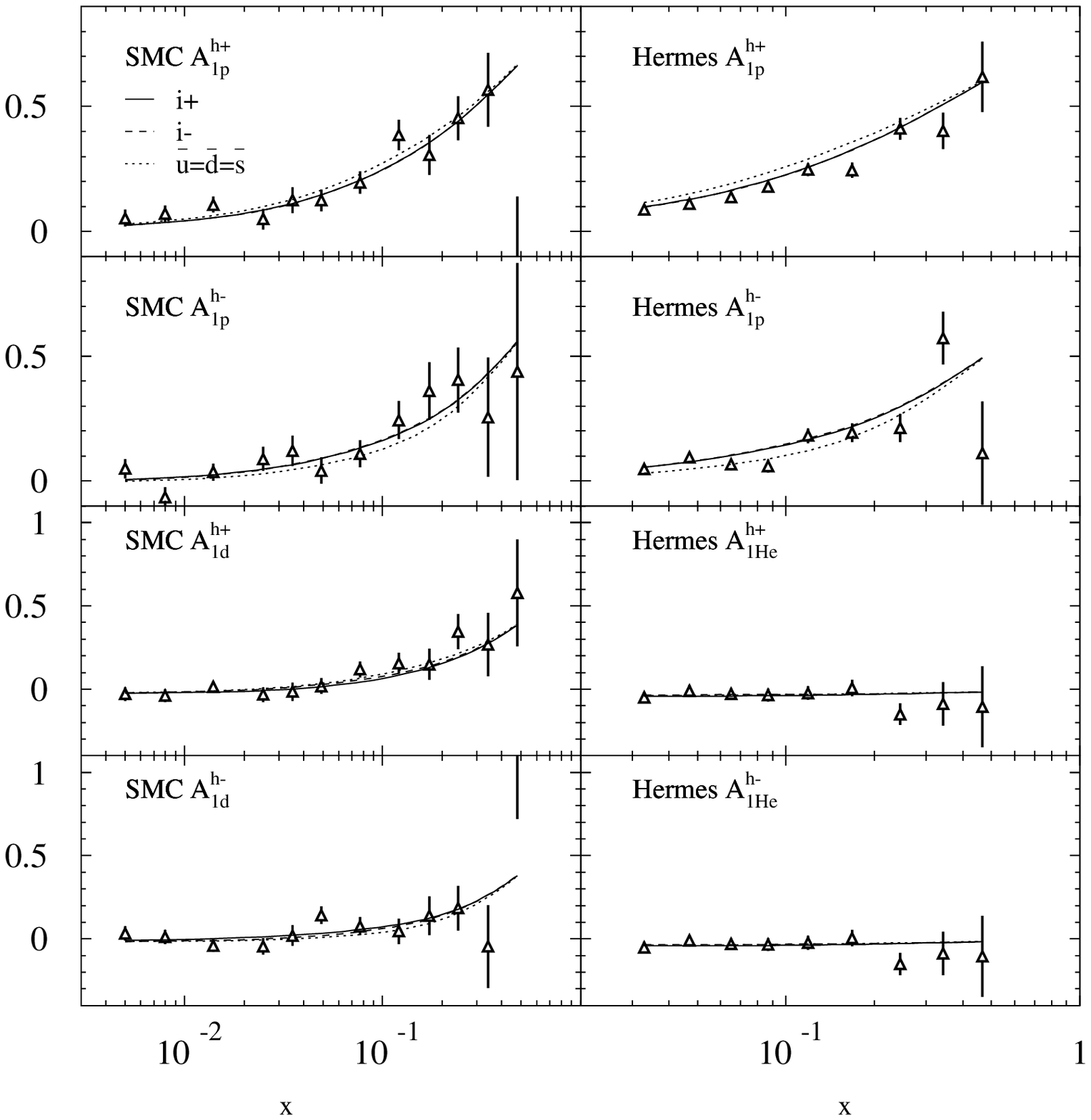}}}
\put(25,5){\mbox{\footnotesize \bf Figure 4:} {\footnotesize Semi-Inclusive
    asymmetries against NLO sets }} 
\end{picture}
\end{figure}

As we mentioned, in the computation of the total $\chi^2$ values we properly take  into account the correlations between SIDIS and inclusive data, finding that its main effect is just to increase slightly the total $\chi^2$ of the fits.

\setlength{\unitlength}{1.mm}
\begin{figure}[!hbt]
\begin{picture}(100,140)(0,0)
\put(5,-20){\mbox{\epsfxsize13.0cm\epsffile{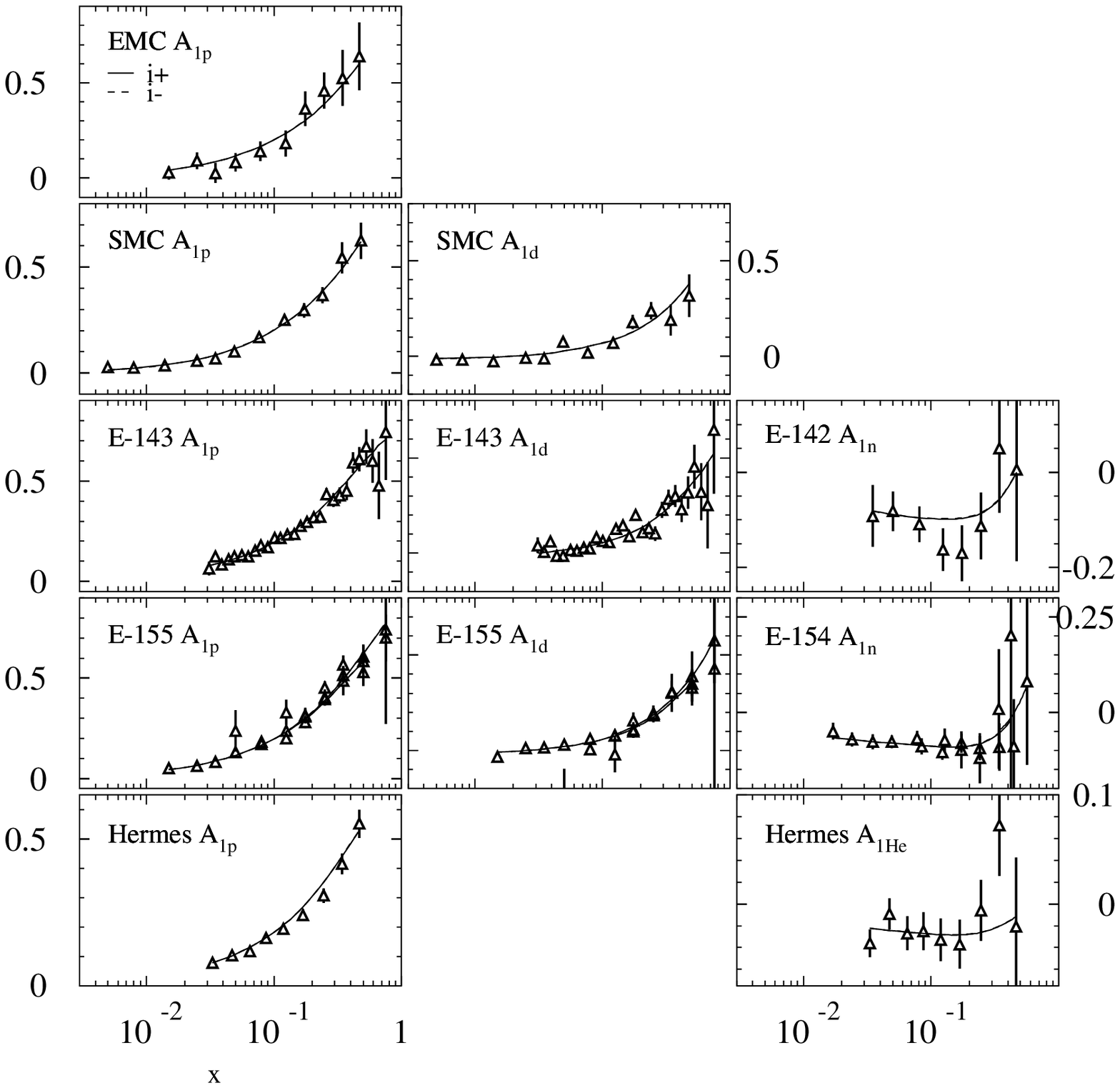}}}
\put(25,-0){\mbox{\footnotesize \bf Figure 5:} {\footnotesize Inclusive asymmetries against NLO sets}} 
\put(5,5){\mbox{    }} 
\end{picture}
\end{figure}

Notice that the six sets of parton densities in Table 4 have positive $\Delta
\overline{u}$ densities, however their sizes are still  correlated to the
choice for  $\Delta \overline{d}$. This is in agreement with the 
discussion below equation (\ref{eqdisc}), since the correlation between 
$\Delta \overline{u}$ and $\Delta \overline{d}$ is such that 
$(\Delta \overline{d} -4 \Delta \overline{u})$ remains roughly the same for
all sets.

 It should be noticed also that even in the case of $SU(2)$-symmetric alternatives (labeled as `$+$'), there are small differences between
the first moment of the $\Delta \overline{u}$ and $\Delta \overline{d}$ at
$Q^2=10$ GeV$^2$, notwithstanding they are set to be equal at the initial scale
$Q_0^2$. The difference comes from the fact that these distributions evolve
differently at NLO due to both the appearance of different non-singlet quark splitting
functions and the non-zero value of $\Delta u_v- \Delta d_v$. Therefore the SU(2) symmetry in the light sea sector is
perturbatively broken, as it happens in the unpolarized case, but the
size of this breakdown is small compared to what is required by the SIDIS data.
Notice that this effect is absent in the LO case, where there is only one $qq$ splitting function.

It is worthwhile mentioning that the slight differences in the $\chi ^2$
between  `$+$' and `$-$' sets are comparable to the uncertainties of the fitting procedure, such as the use of slightly different parameterizations for the initial distributions, the arbitrariness in choosing the initial scale $Q^2_0$, or other external constraints. 
For this reason, we can conclude that $\Delta \overline{d}$ densities are not  well constrained by present data, and the slight differences in $\chi^2$ fail to justify a definite statement about isospin symmetry at sea quark level. 
At variance with $SU(2)$, $SU(3)$ symmetry in the sea  seems to be clearly broken in the sense that
fits prefer $\Delta \overline{u}$  and $\Delta \overline{s}$ with opposite polarization.

In order to try to quantify uncertainties in the spin dependent distributions, specifically, those introduced by the assumptions done about the fragmentation functions, we have varied the parameter $a$ of equation (\ref{eqRD}), which was fixed to 1 in the previous fits, from 0.75 to 1.25. Doing this, we find that this variation produces only negligible changes in the $\Delta \overline{u}$ distribution, which is explained by a large cancellation of its effect in the rate of polarized and unpolarized SIDIS structure functions. Of course, this by no means constitute an exhaustive study of the related uncertainties, but it makes us more confident about the sensitivity of our results to some unavoidable assumption.

Although for the time being we do not intend to perform a comprehensive analysis of other uncertainties involved in the the extraction of the parton distributions presented here, problem which is far from being solved even in the
unpolarized case \cite{errors}, we find reasonable to claim
that present data give a clear signal of a positive $\Delta \overline{u}$
distribution.
Actually, we have found to be impossible to produce parton densities constraining $\Delta \overline{u}$ to be
negative, unless we allow an increase in the $\chi^2$ value of the order of
5\% (about 10-20 units), showing the sensitivity of the semi-inclusive data.

Noticeably, most of the available theoretical calculations \cite{mod} predict
a positive $\Delta \overline{u}$
distribution.
In Figure 6 we show our LO results for $\Delta \overline{u}$ from 
fits  `i$+$' and `i$-$' compared to the prediction of Gl\"uck and Reya, and the one
of Bhalearo \cite{mod} at $Q^2=1$ GeV$^2$. As can be seen, there is a 
reasonable agreement between those predictions and our results in the
kinematical region covered by Hermes semi-inclusive data (i.e. $0.4>x>0.03$),
at least within the expected uncertainties on the extracted distributions. Of
course, there is no experimental evidence yet at smaller values of $x$ so our results just provide an extrapolation in that region. The same models predict also a negative $\Delta \overline{d}$, however as stated before, global fits still do not discriminate a preferred alternative to compare with.  

\setlength{\unitlength}{1.mm}
\begin{figure}[!htb]
\begin{picture}(100,110)(0,0)
\put(22,-7){\mbox{\epsfxsize9.0cm\epsffile{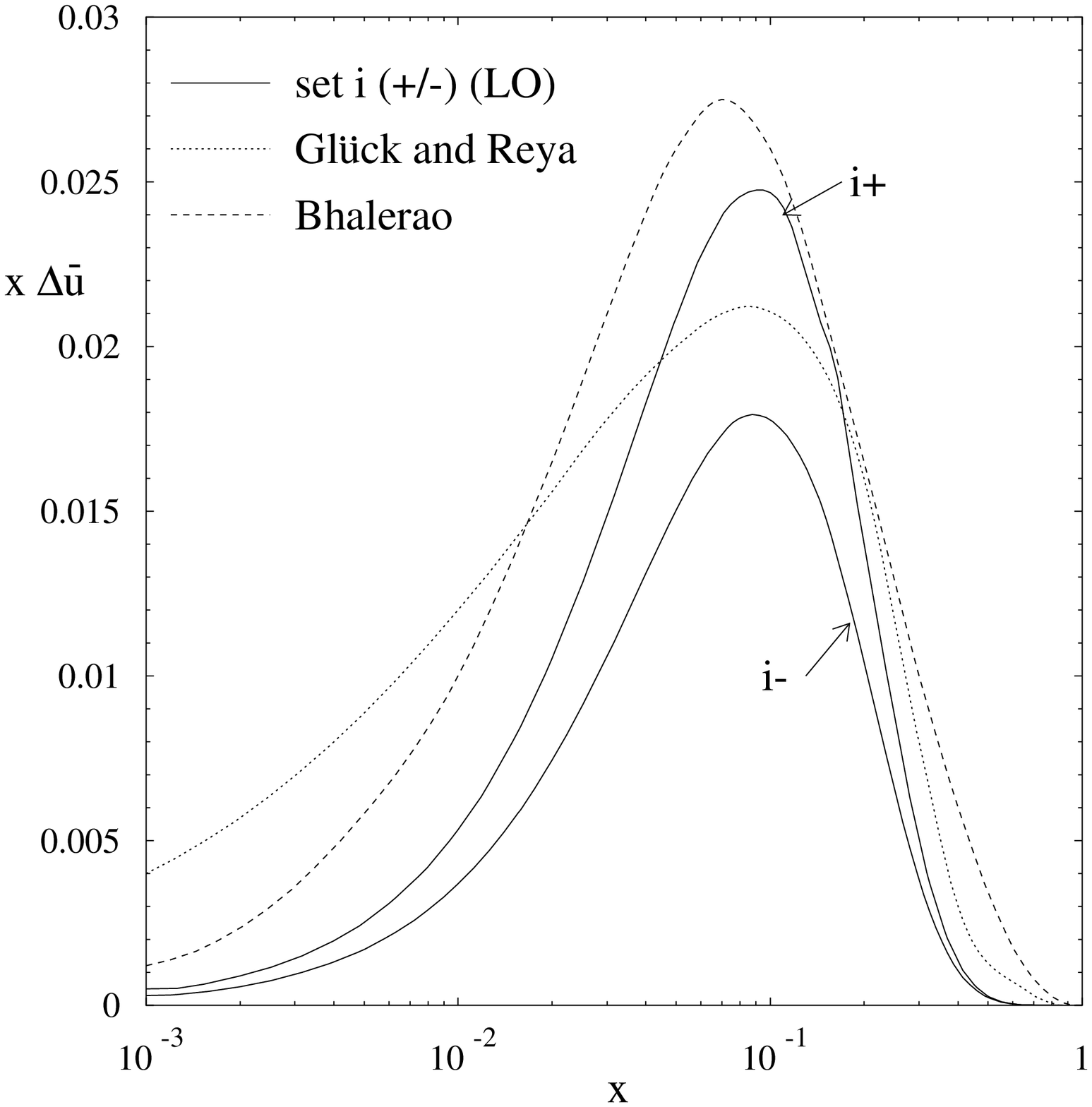}}}
\put(5,0){\mbox{\footnotesize \bf Figure 6:} {\footnotesize  $x \Delta
    \overline{u}$  (set i) at $Q^2=1$ GeV$^2$ compared to 
    the predictions in Ref. \cite{mod}}} 
\end{picture}
\end{figure}

At this point it is worth  comparing our results to the ones reported by the
Hermes Collaboration in reference \cite{Hermes}. There, no definitive sign for $\Delta \overline{u}$ is found. Besides the fact that we take into account the correct $Q^2$ evolution of the distributions both up to LO and NLO accuracy, whereas Hermes uses the naive parton model expressions assuming $Q^2$ independence of the asymmetries, the main difference comes from the fact that 
we perform the fit to {\it all} available inclusive and semi-inclusive data (not only the one coming from Hermes),  and that we do not impose any
constraint  on $\Delta \overline{q}$ distributions. In the Hermes analysis sea
quark distributions are constrained to have the same sign, and taking into
account that the strange quark distribution $\Delta s$ is found to be negative 
already from the inclusive data, this constraint introduce a
strong bias against the $SU(3)$ breaking in the sea as found in our analysis.
\\

\noindent{\Large \bf 5 \,\, Conclusions}\\

The main conclusion of this analysis is that semi-inclusive spin dependent data available today, are not only in perfect agreement with inclusive data with regard to the extraction of  polarized parton densities, but constitute a 
very useful tool to investigate polarization among sea quarks. Specifically,
we have found that SMC and Hermes data show a clear preference for parton
distributions with `up' antiquarks polarized along the nucleon spin and
strange quarks with the opposite polarization \footnote{A Fortran code containing a parameterization of our distributions is available 
upon request} .

Regarding `down' antiquarks, although the global fits  attain marginally lower values of $\chi^2$ when these are polarized in opposition to `up' antiquarks, the differences are so small that we conclude that the data is unable to discriminate between the different alternatives.

The agreement between polarized parton densities obtained from global fits and the constraints coming from hyperon decay data is particularly good in NLO,
with departures of the order of a few percents, but larger for LO sets, partly due to higher order QCD effects.\\

\noindent{\Large \bf Acknowledgments}\\

We thank S. Kretzer for providing us with his routines for fragmentation functions, and A. Br\"ull for kindly providing us Hermes data before publication.

 \pagebreak

\noindent{\Large \bf Appendix: Parameters of the fits}\\

We present here the parameters of the best fits found in our analyses, both at 
LO and NLO. The parameters marked by an asterisk are the ones externally fixed 
 from positivity constraints at large $x$.

{\footnotesize
\begin{center}
\begin{tabular}{|c|c|c|c|c|c|c|} \hline \hline
Parameter  & i +   & i -    & ii +  & ii -   & iii + & iii - \\ \hline
\hline
 $\epsilon_{BJ}$ 
           & -0.021& -0.021 & -0.034& -0.030 &-0.014 & -0.012 \\ \hline
$\epsilon_{SU(3)}$ 
           & -0.002& -0.001 & 0.056 & 0.045 &-0.064 &-0.068 \\ \hline
$\alpha_u$ & 0.873 & 0.877  & 0.921 & 0.918 & 0.936  & 0.933 \\ \hline
$\beta_u^*$& 3.20  &  3.20  & 3.20  & 3.20  & 3.20   & 3.20 \\ \hline
$\gamma_u$ & 14.87 & 14.66  & 13.56 & 13.69 & 14.46  & 14.40 \\ \hline
$\delta_u$ & 1.06  & 1.06   & 1.00  & 1.00  & 0.97  & 0.97 \\ \hline
$\alpha_d$ & 0.460 & 0.460  & 0.474 & 0.465 & 0.446  & 0.443 \\ \hline
$\beta_d^*$& 4.05  & 4.05   & 4.05  & 4.05  & 4.05   & 4.05 \\ \hline
$\gamma_d$ & 14.99 & 14.99  & 14.81 & 14.99 & 14.99 & 14.99 \\ \hline
$\delta_d$ & 1.65  & 1.67   & 1.66  & 1.66  & 1.62   & 1.63 \\ \hline
$N_s$    & -0.062 &-0.062 &-0.069&-0.069& -0.063& -0.063 \\ \hline
$\alpha_s$ & 2.49  & 2.49   & 2.50  & 2.50  & 2.49   & 2.49 \\ \hline
$\beta_s$ & 10.00  & 10.00  & 10.16 & 10.16 & 10.32  & 10.32 \\ \hline
$N_{\bar{u}}$& 0.065&0.046& 0.059& 0.044& 0.055 & 0.042 \\ \hline
$\alpha_{\bar{u}}$&1.01&1.12& 1.25& 1.15 & 1.12  & 1.17 \\ \hline
$N_g$      & 0.25 & 0.27  & 0.40 & 0.40 & 0.7   & 0.7 \\ \hline
$\alpha_g$ & 1.5 & 1.5      & 1.5   & 1.5 & 1.5  & 1.5 \\ \hline
$\beta_g^*$ & 6.0 & 6.0 & 6.0 & 6.0 & 6.0   & 6.0  \\ \hline
\hline
$\chi^2$ &  330.52 & 330.00 & 330.98& 330.22& 332.31 & 331.64\\ \hline
\hline \end{tabular}
\end{center}}
\begin{center}
{\footnotesize {\bf Table A1}: Coefficients for NLO total sets}
\end{center}
\pagebreak

{\footnotesize
\begin{center}
\begin{tabular}{|c|c|c|c|c|c|c|} \hline \hline
Parameter  & i + & i - & ii + & ii - & iii + & iii - \\ \hline \hline
$\epsilon_{BJ}$ 
           & -0.131& -0.131 & -0.133 & -0.135 & -0.14 & -0.136 \\ \hline
$\epsilon_{SU(3)}$ 
           & 0.071 & 0.067 & 0.032 & 0.047 & 0.036 & 0.039 \\ \hline

$\alpha_u$ & 0.768 & 0.777 & 0.766 & 0.768 & 0.77 & 0.767 \\ \hline

$\beta_u^*$ & 3.05 &  3.05 & 3.05 & 3.05 & 3.05 & 3.05 \\ \hline

$\gamma_u$ & 14.88 & 14.88 & 14.97 & 14.81 & 14.98 & 14.99 \\ \hline

$\delta_u$ & 1.64  & 1.64 & 1.69 & 1.69 & 1.69 & 1.69 \\ \hline

$\alpha_d$ & 0.5   & 0.499 & 0.502 & 0.52 & 0.532 & 0.527 \\ \hline

$\beta_d^*$ & 3.77 & 3.77 & 3.77 & 3.77 & 3.77 & 3.77 \\ \hline

$\gamma_d$ & 14.55 & 14.99 & 15.0 & 13.24 & 13.76 & 13.98 \\ \hline

$\delta_d$ & 1.68  & 1.75 & 1.69 & 1.69 & 1.69 & 1.69 \\ \hline

$N_s$     & -0.069 & -0.07 & -0.068 & -0.070 & -0.069 & -0.069 \\ \hline

$\alpha_s$ & 2.37  & 2.4 & 2.19 & 2.25 & 2.30 & 2.32 \\ \hline

$\beta_s$ & 11.42  & 11.42 & 11.42 & 11.42 & 11.42 & 11.47 \\ \hline

$N_{\bar{u}}$&0.060& 0.043 & 0.058 & 0.042 & 0.056 & 0.041 \\ \hline

$\alpha_{\bar{u}}$ &  1.19 & 1.19 & 1.2 & 1.2 & 1.2 & 1.21 \\ \hline

$N_g$ & 0.35 & 0.35& 0.6 & 0.6 & 0.7 & 0.7 \\ \hline

$\alpha_g$ & 1.8   & 1.78 & 1.8 & 1.8 & 1.8 & 1.8 \\ \hline

$\beta_g^*$ & 6.0  & 6.0 & 6.0 & 6.0 & 6.0 & 6.0  \\ \hline
\hline

$\chi^2$ &  336.24 & 335.54 & 334.19 &  334.03 & 334.26 & 333.69 \\ \hline

\hline \hline
\end{tabular}
\end{center}}
\begin{center}
{\footnotesize {\bf Table A2}: Coefficients for LO total sets}
\end{center}

{\footnotesize
\begin{center}
\begin{tabular}{|c|c|c|c|c|c|c|} \hline \hline
           & \multicolumn{3}{|c|}{NLO} & \multicolumn{3}{|c|}{LO} \\ \hline \hline

Parameter  & i  & ii & iii & i  & ii & iii \\ \hline
 \hline
$\epsilon_{BJ}$ & -0.015 & -0.018 & -0.0025 & -0.126 & -0.126 & -0.126 \\ \hline
$\epsilon_{SU(3)}$ & -0.003 & 0.037 & -0.063 & 0.016 & 0.018 & 0.009 \\ \hline
$\alpha_u$ & 0.876 & 0.915 & 0.928& 0.766 & 0.77 & 0.771 \\ \hline
$\beta_u^*$ & 3.2 & 3.2 & 3.2 & 3.05 & 3.05 & 3.05 \\ \hline
$\gamma_u$ & 14.53 & 13.37  & 14.38& 14.99 & 14.94  & 14.84  \\ \hline
$\delta_u$ & 1.06 & 1.0 & 0.97 & 1.66 & 1.68 & 1.68 \\ \hline
$\alpha_d$ & 0.46 & 0.47 & 0.43& 0.471 & 0.477 & 0.49  \\ \hline
$\beta_d^*$ & 4.05 & 4.05 & 4.05 & 3.77 & 3.77 & 3.77 \\ \hline
$\gamma_d$ & 14.99 & 14.85 & 14.66 & 15.0 & 14.98 & 14.97 \\ \hline
$\delta_d$ & 1.675 & 1.66 & 1.55& 1.72 & 1.68 & 1.68 \\ \hline
$N_s$  & -0.062 & -0.069 & -0.064 & -0.065 & -0.068 & -0.069 \\ \hline
$\alpha_s$ & 2.49 & 2.5 & 2.49 & 2.38 & 2.38 & 2.38 \\ \hline
$\beta_s$ & 10.0 & 10.16 & 10.32 & 11.42 & 11.42 & 11.42 \\ \hline
$N_g$ & 0.25 & 0.4 & 0.7 & 0.35 & 0.59 & 0.7  \\ \hline
$\alpha_g$ & 1.5 & 1.5 & 1.5& 1.8 & 1.8 & 1.75  \\ \hline
$\beta_g^*$ & 6.0 & 6.0 & 6.0 & 6.0 & 6.0 & 6.0  \\ \hline
\hline
$\chi^2$ &  228.66 & 228.69 & 230.33&  237.55 & 236.1 & 236.44 \\ \hline

\hline \hline
\end{tabular}
\end{center}}
\begin{center}
{\footnotesize {\bf Table A3}: Coefficients for LO and NLO inclusive sets}
\end{center}

\end{document}